\documentclass{article}

\usepackage{amsfonts}
\usepackage{amsmath}
\usepackage{amssymb}
\usepackage{graphicx}

\def\dual{\overline}
\newcommand{\CHAIN}[1]{\mathbf{#1}}
\newcommand{\move}[1]{#1}
\def\P{{\CHAIN{P}}}
\def\Q{{\CHAIN{Q}}}

\begin{document}

\title{Inferences about Interactions:\\
Fermions and the Dirac Equation}

%\classification{03.65.Pm, 02.50.Cw, 03.30.+p, 03.67.-a}
%\keywords      {Dirac equation, Fermion physics, probability theory, relativity, quantum mechanics, inference, quantification, Zitterbewegung, zitter}

\author{Kevin H. Knuth\\ Departments of Physics and Informatics\\
University at Albany (SUNY)\\ Albany NY 12222, USA}

\date{\today}
\maketitle

\begin{abstract}
At a fundamental level every measurement process relies on an interaction where one entity influences another.  The boundary of an interaction is given by a pair of events, which can be ordered by virtue of the interaction.   This results in a partially ordered set (poset) of events often referred to as a causal set.  In this framework, an observer can be represented by a chain of events.  Quantification of events and pairs of events, referred to as intervals, can be performed by projecting them onto an observer chain, or even a pair of observer chains, which in specific situations leads to a Minkowski metric replete with Lorentz transformations.  We illustrate how this framework of interaction events gives rise to some of the well-known properties of the Fermions, such as Zitterbewegung.  We then take this further by making inferences about events, which is performed by employing the process calculus, which coincides with the Feynman path integral formulation of quantum mechanics.  We show that in the 1+1 dimensional case this results in the Feynman checkerboard model of the Dirac equation describing a Fermion at rest.
\end{abstract}

%\maketitle

%%%%%%%%%%%%%%%%%%%%%%%%%%%%%%%%%%%%%%%%%%%%
%% MAINMATTER
%%%%%%%%%%%%%%%%%%%%%%%%%%%%%%%%%%%%%%%%%%%%

\section{Introduction}

This paper summarizes the synthesis of three distinct research threads which were inspired by the quantification of Boolean algebra by Cox \cite{Cox:1946}.  The first represents the current perspective of the quantification of the Boolean lattice of logical statements, which results in probability theory \cite{Knuth:measuring, Knuth&Skilling:2012}.  The second is the pair-wise quantification of the algebra of combining measurement sequence outcomes.  Combining this with probability theory results in the process calculus, which represents the application of inference to quantum mechanical measurement sequences \cite{GKS:PRA,GK:Symmetry} and has been shown to be in accord with the Feynman path integral formulation of quantum mechanics \cite{Feynman:1948}.  The third is the quantification of partially ordered sets via chain projection, which recovers the fundamental characteristics of space-time physics \cite{Knuth+Bahreyni:EventPhysics}.    We show that these theories applied together enable one to make inferences about a simple model of direct particle-particle interaction in a way that reproduces well-known properties unique to Fermion physics---including the Dirac equation.

We assume that the reader is familiar with Cox's derivation of probability theory \cite{Cox:1946}, as well as more recent work extending these ideas to distributive lattices \cite{Knuth:measuring, Knuth&Skilling:2012}.  Below we summarize the other two research threads, and proceed to apply them to our model.

\subsection{Process Calculus and Quantum Mechanical Inference}

We begin by summarizing the results from the derivation of the process calculus and its relation to quantum mechanical inference.  The interested reader is referred to the following papers on probability theory \cite{Knuth&Skilling:2012} and quantum mechanical sequences \cite{GKS:PRA,GK:Symmetry} for more detailed information.

In a run of an experiment, a physical system originating from a source passes through a sequence of measurements $M_1, M_2, \ldots$, which produce outcomes $m_1,m_2, \ldots$, respectively. We summarize these outcomes by denoting a measurement sequence $[m_1,m_2, \ldots]$.  Some measurements may be coarse-grained such that they do not give definitive results, but rather generate outcomes consisting of a set of results.  For example, imagine that we perform three measurements $M_1, M_2$, and $M3$ in succession such that $M_1$ generates the outcome $m_1$, $m_2$ generates the coarse-grained result $\{m_2, m_2'\}$, and $M3$ generates the outcome $m_3$.  We denote this sequence as $[m_1, (m_2,m_2'), m_3]$.

We can conceive of the coarse-grained measurement $M_2$ as performing two measurements in parallel.  This enables one to introduce the \textbf{\emph{parallel operation}} $\vee$ which combines the sequences associated with two separate experiments to obtain the sequence corresponding to the coarse grained measurement above
\begin{equation}
[m_1, (m_2,m_2'), m_3] = [m_1, m_2, m_3] \vee [m_1, m_2', m_3].
\end{equation}

Similarly, we can conceive of the experiment comprised of three successive measurements $M_1, M_2, M_3$ in terms of a concatenation of two separate experiments, the first consisting of measurements $M_1, M_2$ yielding outcomes $m_1, m_2$ and the second consisting of measurements $M_2, M_3$ yielding outcomes $m_2, m_3$.  We can express this as a \textbf{\emph{series operation}} acting on the measurement outcomes
\begin{equation}
[m_1, m_2, m_3] = [m_1, m_2] \cdot [m_2, m_3].
\end{equation}

The parallel operation is both commutative and associative; whereas, the series operation is only associative. In addition, the algebraic properties of these operators includes both right- and left-distributivity of $\cdot$ over $\vee$.  These algebraic properties constrain quantification of the measurement sequences by real-valued pairs.  Associativity of the parallel operator results in a sum rule for the pair so that for a sequence $[m_1, m_2, m_3]$ quantified by $(a_1,a_2)$ and a sequence $[m_1, m_2', m_3]$ quantified by $(b_1,b_2)$ we have that the sequence $[m_1, (m_2,m_2'), m_3]$ is quantified by
\begin{equation}
(a_1, a_2) \oplus (b_1, b_2) = (a_1+b_1, a_2+b_2).
\end{equation}
Distributivity together with agreement with probability theory results in a product rule, which enables one to view the pair as a complex number.  For example, for a sequence $[m_1, m_2]$ quantified by $(a_1,a_2)$ and a sequence $[m_2, m_3]$ quantified by $(b_1,b_2)$ we have that the sequence $[m_1, m_2, m_3]$ is quantified by
\begin{equation}
(a_1, a_2) \odot (b_1, b_2) = (a_1b_1 - a_2b_2, a_1b_2 + a_2b_1).
\end{equation}
Last, agreement with probability theory results in the \emph{\textbf{Born Rule}}, which maps the pair $(a_1,a_2)$ quantifying $[m_1, m_2, m_3]$ to the probability $Prob(m_2, m_3 | m_1)$ by
\begin{equation}
Prob(m_2, m_3 | m_1) = a_1^2 + a_2^2,
\end{equation}
which can be written in complex form as $a^{*}a$ where $a = a_1 + ia_2$.  Given the fact that this is identical to the Feynman path integral formulation of quantum mechanics, we refer to the quantifying pair as the amplitude, and note that the modulus squared gives the probability.

\subsection{Chain Projection and Space-time Physics}

In this section we summarize the mathematics of poset quantification by chain projection and its relation to events and interactions \cite{Knuth+Bahreyni:EventPhysics}, which we call the \emph{\textbf{poset picture}}. Later we will relate this picture to a different perspective, which we call the \emph{\textbf{space-time picture}}.

We conceive of the \emph{boundary} of an interaction as a pair of causally-related events, where one event represents the act of influencing and the other event represents the act of being influenced.  In a system consisting of interacting particles (for lack of a better word), this results in a partially ordered set (poset) of events.  Quantification proceeds by distinguishing an observer chain $\P$ consisting of a set of totally ordered elements and assigning each element $p_i$ a real number $v_{\P}(p_i)$ such that $v_{\P}(p_x) \leq v_{\P}(p_y)$ for $p_x \leq p_y$.

\begin{figure}[t]
\centering
  \includegraphics[height=0.33\textheight]{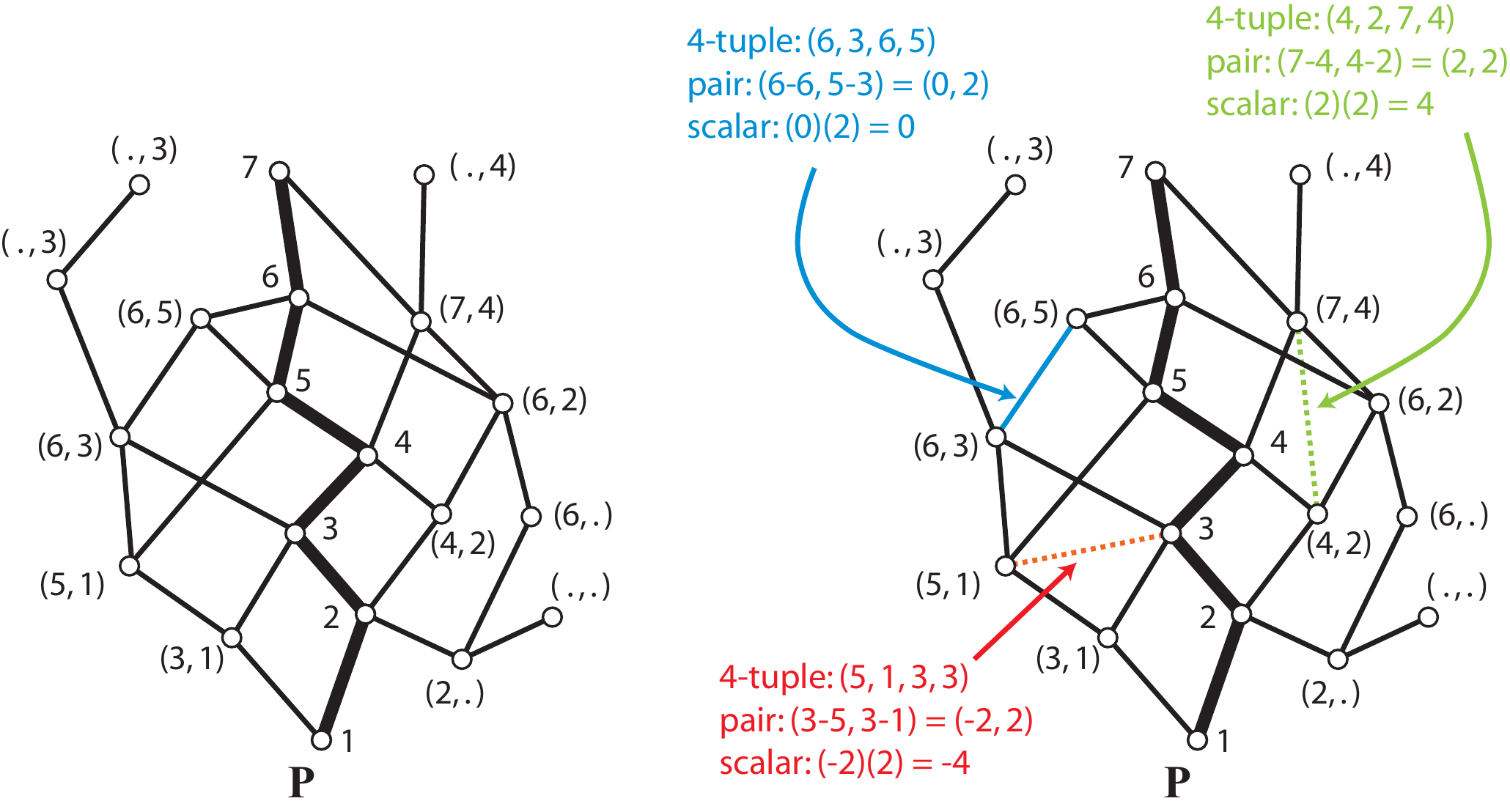}
  \caption{(Left) This figure illustrates a poset with elements quantified by a chain $\P$.  Note that not all elements project onto $\P$ so as to be quantified by two numbers.  In fact, some elements do not project onto $\P$ at all (which is represented by using a dot in the pair). (Right) Here we illustrate the quantification of intervals (pairs of poset elements).  Intervals can be quantified with four numbers (4-tuple), two numbers (interval pair) or a single number (interval scalar).  The interval scalar enables one to distinguish three classes of intervals: chain-like (positive interval scalar), antichain-like (negative interval scalar), or projection-like (interval scalar equals zero).  In the space-time picture, these three classes of intervals are called time-like, space-like and light-like, respectively.}
  \label{fig:quantified-poset}
\end{figure}

Some events in the poset will both include an element of the chain $\P$ and be included by an element of the chain $\P$.  For such an event $x$ we can identify the least element $Px$ of the chain that includes the event, as well as the greatest element $\dual{P}x$ of the chain that is included by the event.  If these elements exist, we can treat these two maps in terms projection operators, $P$ and $\dual{P}$ taking an element of the poset to an element on the chain.  This allows us to quantify the poset element $x$ with the pair $(v_{\P}(Px), v_{\P}(\dual{P}x))$ as illustrated in Figure \ref{fig:quantified-poset} (Left).

Two poset elements $x$ and $y$ can be compared by projecting both onto the chain and collecting the resulting pair into a 4-tuple
$$
(v_{\P}(Px), v_{\P}(\dual{P}x), v_{\P}(Py), v_{\P}(\dual{P}y)).
$$
We define the interval $[x,y]$ in terms of the poset elements $x$ and $y$ and note that this interval projects to two intervals on $\P$, $[Px, Py]$ and $[\dual{P}x, \dual{P}y]$, with respective \textbf{\emph{lengths}} $v_{\P}(Py) - v_{\P}(Px)$ and $v_{\P}(\dual{P}y) - v_{\P}(\dual{P}x)$.  This enables us two also quantify the interval with a pair
$$
(\Delta P, \Delta \dual{P}) = (v_{\P}(Py) - v_{\P}(Px), v_{\P}(\dual{P}y) - v_{\P}(\dual{P}x)).
$$

We now consider the case where there exists a chain $\Q$ such that $\P$ and $\Q$ agree in the lengths of intervals $[p_i, p_j] \subseteq [p_{min}, p_{max}] \subseteq \P$ and $[q_i, q_j] \subseteq [q_{min}, q_{max}] \subseteq \Q$.  This is equivalent to assuming a locally flat space-time where $\P$ and $\Q$ define an inertial frame.  We also consider that the two elements $x$ and $y$ are situated \textbf{\emph{between}} $\P$ and $\Q$, which is defined by requiring that $Qx = Q\dual{P}x$ and $Px = P\dual{Q}x$ such that $Px, Py \in [p_{min}, p_{max}]$ and $Qx, Qy \in [q_{min}, q_{max}]$. Then we can write the \textbf{\emph{interval pair}} above as
\begin{equation} \label{eq:interval-pair}
(\Delta p, \Delta q) = (v_{\P}(Py) - v_{\P}(Px), v_{\Q}(Qy) - v_{\Q}(Qx)).
\end{equation}

One can show that from this pair, one can obtain a unique scalar measure,
\begin{equation}
\Delta p \Delta q = (v_{\P}(Py) - v_{\P}(Px))(v_{\Q}(Qy) - v_{\Q}(Qx))
\end{equation}
which we call the \textbf{\emph{interval scalar}}.  The interval scalar enables one to distinguish three classes of intervals: chain-like (positive interval scalar), antichain-like (negative interval scalar), or projection-like (interval scalar equals zero).  These three measures are illustrated on three intervals highlighted in Figure \ref{fig:quantified-poset} (Right).

The \emph{\textbf{space-time picture}} is attained by recognizing that in general an interval can be decomposed into the concatenation of two intervals whose interval pairs are symmetric and antisymmetric and sum according to
\begin{equation}
(\Delta p, \Delta q) = \Big(\frac{\Delta p + \Delta q}{2} , \frac{\Delta p + \Delta q}{2} \Big) + \Big(\frac{\Delta p - \Delta q}{2} , \frac{\Delta q - \Delta p}{2} \Big).
\end{equation}
Furthermore, the interval scalar decomposes similarly
\begin{equation}
\Delta p \Delta q = \Big(\frac{\Delta p + \Delta q}{2}\Big)^2 - \Big(\frac{\Delta p - \Delta q}{2}\Big)^2.
\end{equation}

This suggests a convenient change of variables where
\begin{equation}
\Delta t = \frac{\Delta p + \Delta q}{2}
\end{equation}
and
\begin{equation}
\Delta x = \frac{\Delta p - \Delta q}{2}.
\end{equation}
The pair can then be written as
\begin{equation}
(\Delta p, \Delta q) = (\Delta t, \Delta t) + (\Delta x, -\Delta x)
\end{equation}
and the scalar as
\begin{equation}
\Delta p \Delta q = {\Delta t}^2 - {\Delta x}^2,
\end{equation}
which is the Minkowski metric.

By studying how the interval pair transforms as one changes the quantification basis from one pair of chains to another pair of chains, linearly-related to the first, one finds the relevant variable to be
\begin{equation} \label{eq:speed}
\beta = \frac{\Delta p - \Delta q}{\Delta p + \Delta q} = \frac{\Delta x}{\Delta t},
\end{equation}
which is the poset analogue of speed.  Keep in mind that nothing moves in the poset picture.  Instead what we call motion in the space-time picture is a representation of interaction in the poset picture.

Space-time emerges from the characterization of the partially ordered set by a set of distinguished embedded chains.

\section{Direct Particle-Particle Interaction}

\begin{figure}[t]
\centering
  \includegraphics[height=0.25\textheight]{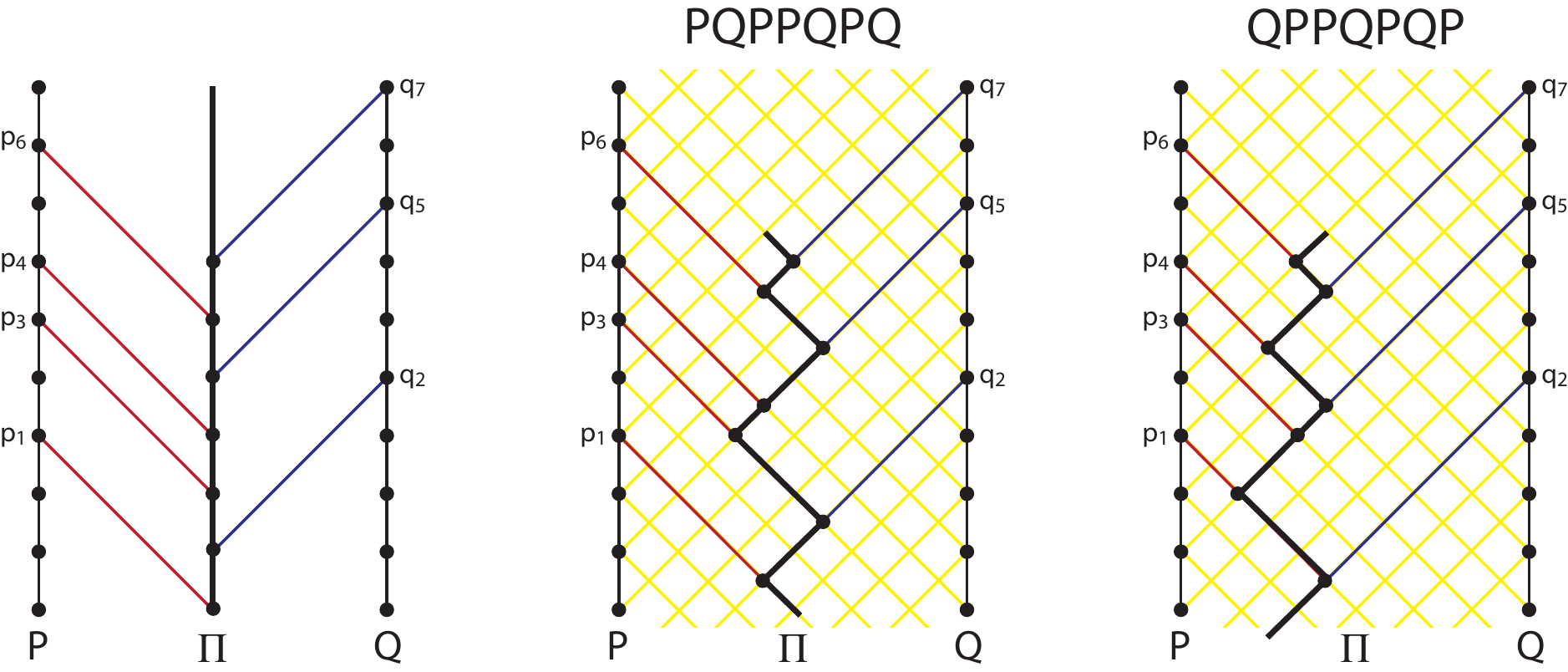}
  \caption{This figure illustrates the poset representing a free particle.  (Left) The standard Haase diagram for posets illustrating the connectivity.  (Center) Here the connectivity is displayed on a space-time diagram (time progressing upward). The particle chain has been drawn to illustrate the sequence (PQPPQPQ), which is only one of the many possible space-time paths it can be interpreted as to having taken.  Note that the particle is interpreted as zig-zagging at the speed of light (Zitterbewegung) with \move{P}-moves being represented by a diagonal move to the upper-right and \move{Q}-moves being represented by a diagonal move to the upper-left as if it were moving on a checkerboard. (Right) The particle chain has been drawn to illustrate the sequence (QPPQPQP).}
  \label{fig:free-particle}
\end{figure}

We consider direct particle-particle interactions in the 1+1-dimensional case of the \emph{\textbf{free particle}}, by which we mean that the particle can influence other particles, but is not itself influenced.  We model this by a chain of events $\Pi$, that influences its surroundings, but is not influenced itself.  This free particle is assumed to be situated in between two coordinated observer chains $\P$ and $\Q$ that define a 1+1 dimensional locally flat space-time, as described by the conditions leading to (\ref{eq:interval-pair}) above and illustrated in Figure \ref{fig:free-particle}.

The interactions themselves are not directly observed, but instead only the effects of the interactions are noted as $\P$ and $\Q$ are influenced repeatedly by $\Pi$.  The result is a series of detections $p_1, p_3, p_4, p_6$ made by $\P$ and a series of detections $q_2, q_5, q_7$ made by $\Q$.  While these detections are ordered along the respective chains, there is not enough information for the observers to collectively perform a reconstruction of the interaction pattern, or connectivity, in the poset picture, which is analogous to a tomographic reconstruction of the path of particle $\Pi$ in space-time.  The result is that neither the particle's position nor speed can be precisely specified.

More specifically, for four interactions with $\P$ and and three interactions with $\Q$, there exist ${{4+3} \choose 4} = {{4+3} \choose 3} = 35$ possible interaction sequences (space-time paths) that could have given rise to the observations made by $\P$ and $\Q$.
These include:
\[ \begin{array}{llll}
PPPPQQQ & PQPPPQQ & QPPPPQQ & QQPPPPQ \\
PPPQPQQ & PQPPQPQ & QPPPQPQ & QQPPPQP \\
PPPQQPQ & PQPQPPQ & QPPQPPQ & QQPPQPP \\
PPPQQQP & PQPPQQP & QPPPQQP & QQPQPPP \\
PPQPPQQ & PQPQPQP & QPPQPQP & QQQPPPP \\
PPQPQPQ & PQPQQPP & QPPQQPP &         \\
PPQQPPQ & PQQPPPQ & QPQPPPQ &         \\
PPQPQQP & PQQPPQP & QPQPPQP &         \\
PPQQPQP & PQQPQPP & QPQPQPP &         \\
PPQQQPP & PQQQPPP & QPQQPPP & \end{array}\]

These sequences represent the set of possible bit strings (eg. $P=0$, $Q=1$) constructed from the detections (\emph{\textbf{bit from it}}) from which we can make inferences (\emph{\textbf{it from bit}}) about the particle's behavior.  In the space-time picture, these bit strings represent all the possible paths that the particle could have taken to get from the initial to final states.

Let us consider how the observer chains interpret the particle's behavior.  We can either consider the correct unknown sequence, or any one of the hypothesized sequences reconstructed from the observations.  By projecting intervals defined by successive interactions along the particle chain $\Pi$ onto the observer chains $\P$ and $\Q$ we see that one component of the interval pair is always zero so that intervals along the particle chain are always projection-like, or light-like.  That is, the particle always undertakes bishop-moves as if on a chessboard with the direction of the move attributed to the interval being dictated by the direction of the interaction associated with the first event defining the interval.  Thus the particle is interpreted in the space-time picture as zig-zagging back and forth at speeds, given by (\ref{eq:speed}), of $\beta = \pm 1$, which is the speed of light.  This behavior, known as \emph{\textbf{Zitterbewegung}}, or \emph{\textbf{zitter}} for short, which is German for `trembling motion', was first proposed by Schrodinger who found that the Dirac equation only admits velocity eigenvalues of $\pm 1$ indicating that Fermions can \emph{only} move at the speed of light.

This behavior is reproduced here as a consequence of direct particle-particle interaction.  Interactions involving only pairs of particles in the poset picture result in particles being able to move only at the speed of light in the space-time picture.  Finite subluminal speeds are attained only on average.  Thus the very same interactions that set up what is interpreted as space-time are responsible for a particle's trembling motion, which is related to inertia as this behavior prevents massive particles from traveling at the speed of light.

\section{Dirac 1+1}

Here we briefly illustrate how the Dirac equation arises from this simple model.  We consider the case of a particle at rest. The analysis begins by deriving the assignment of amplitudes to measurement sequences, and concludes by describing how this framework leads directly to the Feynman checkerboard model, which is known to result in the Dirac equation in 1+1 dimensions.

\subsection{Deriving Amplitudes}

\begin{figure}[t]
\centering
  \includegraphics[height=0.25\textheight]{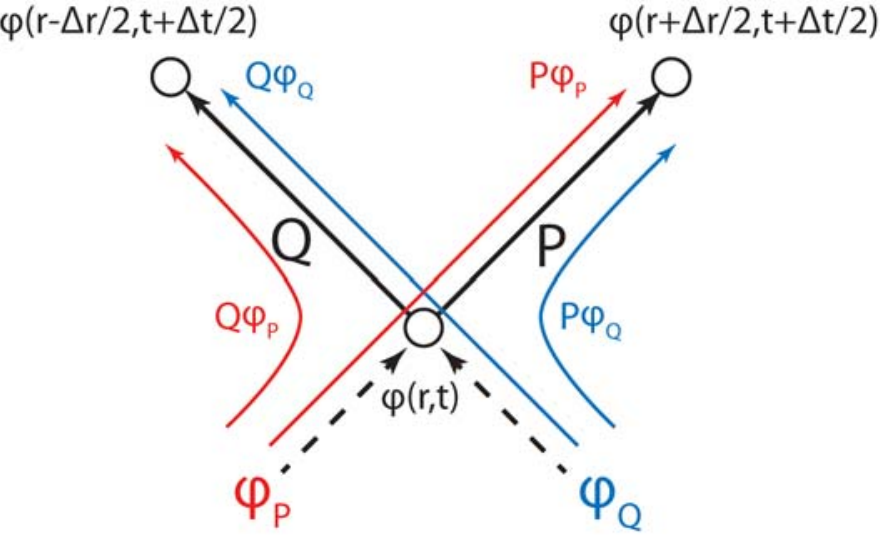}
  \caption{This figure illustrates the four amplitudes that go into relating $\phi(r,t)$ to either $\phi(r-\frac{\Delta}{2},t+\frac{\Delta}{2})$ or $\phi(r+\frac{\Delta}{2},t+\frac{\Delta}{2})$, which occurs with probability one.  There are two possible ways in which the initial state can occur with the particle previously having undergone either a \move{P-move} or a \move{Q-move}.}
  \label{fig:two-component-spinor-p1}
\end{figure}

A free particle's direct particle-particle interaction can be viewed as a single transition from one of two possible initial sequence states, \move{P} or \move{Q}, to the two possible subsequence states \move{P} and \move{Q} as illustrated in Figure \ref{fig:two-component-spinor-p1}.  Such a transition is represented in the space-time picture as a particle experiencing a transition from some initial state given by $(r,t)$ to one of two possible final states: $(r-\frac{\Delta r}{2}, t+\frac{\Delta t}{2})$ or $(r+\frac{\Delta r}{2}, t+\frac{\Delta t}{2})$. Since this must occur with probability one, we can write
\begin{equation}  \label{eq:prob-one-transition}
Prob\Big((r-\frac{\Delta r}{2}, t+\frac{\Delta t}{2}) | (r,t) \Big) + Prob\Big((r+\frac{\Delta r}{2}, t+\frac{\Delta t}{2}) | (r,t) \Big) = 1.
\end{equation}

The initial state is quantified by a pair of amplitudes (a two-component complex-valued quantity reminiscent of a spinor) collectively written as $\phi(r,t) \equiv \phi$, which represent the amplitude $\phi_P$ of a particle having the initial sequence state \move{P} and the amplitude $\phi_Q$ of a particle having the initial sequence state \move{Q}.  We write this as
\begin{equation}
\phi(r,t) \equiv \phi = \left(\begin{array}{c} \phi_P\\ \phi_Q \end{array}\right).
\end{equation}
A \move{P-move} multiplies each of these two possible states by one of two amplitudes, as does a \move{Q-move}.  We can conveniently represent this as a matrix multiplication with matrices $P$ and $Q$, though it is important to remember that the two-component `spinor' is merely keeping track of two possible initial states and the matrix is propagating these states independently using the product rule of the process calculus.

The probability can be calculated via the Born rule so that (\ref{eq:prob-one-transition}) can be written compactly as
\begin{equation}
(Q \phi)^\dagger (Q \phi) + (P \phi)^\dagger (P \phi) = 1.
\end{equation}
This can be rewritten as
\begin{equation}
\phi^\dagger (Q^\dagger Q + P^\dagger P) \phi = 1,
\end{equation}
which implies that
\begin{equation} \label{eq:PQ-condition}
Q^\dagger Q + P^\dagger P = I
\end{equation}
where $I$ is the identity matrix.

Since the matrix $P$ takes the amplitude for either a \move{P} or \move{Q} move and appends (via complex multiplication of the amplitudes) a \move{P} move, the matrix $P$ is of the form
\begin{equation}
P = \begin{pmatrix}
x&y \\
0&0 \end{pmatrix}.
\end{equation}
For a particle at rest, the \move{P} and \move{Q} moves are symmetric.  Thus $Q$ is of the form
\begin{equation}
Q = \begin{pmatrix}
0&0 \\
y&x \end{pmatrix}
\end{equation}
where the non-zero components of $Q$ are related to those of $P$ by the symmetry of the projections onto the chains $\CHAIN{P}$ and $\CHAIN{Q}$.

Condition (\ref{eq:PQ-condition}) can be rewritten as
\begin{equation}
\begin{pmatrix}
0 & y^* \\
0 & x^* \end{pmatrix}
\begin{pmatrix}
0 & 0 \\
y & x \end{pmatrix}
+
\begin{pmatrix}
x^* & 0 \\
y^* & 0 \end{pmatrix}
\begin{pmatrix}
x & y \\
0 & 0 \end{pmatrix}
=
\begin{pmatrix}
1 & 0 \\
0 & 1 \end{pmatrix}
\end{equation}
which becomes
\begin{equation}
\begin{pmatrix}
y^*y & y^*x \\
x^*y & x^*x \end{pmatrix}
+
\begin{pmatrix}
x^*x & x^*y \\
y^*x & y*y \end{pmatrix}
=
\begin{pmatrix}
1 & 0 \\
0 & 1 \end{pmatrix}
\end{equation}
resulting in two conditions
\begin{equation} \label{eq:complex-norm-condition}
x^{*} x + y^{*} y = 1
\end{equation}
\begin{equation} \label{eq:complex-off-diag-condition}
x^{*} y = -y^{*} x.
\end{equation}

Writing $x = a e^{i\alpha}$ and $y = b e^{i\beta}$, we have from (\ref{eq:complex-norm-condition}) that
\begin{equation}
a^2 + b^2 = 1.
\end{equation}
Since there are two unknowns, we have some freedom to assign amplitudes.  The parameters $a$ and $b$ determine the relative probability that one type of move will be followed by another thus controlling the degree to which the particle zig-zags.  We hypothesize that this is related to the mass of the particle, and in this treatment we choose a particular mass by setting $a = b$ so that
\begin{equation}
a = b = \frac{1}{\sqrt{2}}.
\end{equation}

From (\ref{eq:complex-off-diag-condition}), we have that
\begin{equation}
e^{i\theta} + e^{-i\theta} = 0,
\end{equation}
where $\theta = \alpha - \beta$ is a relative phase, which is then found to be $\theta \in \{ \frac{\pi}{2}, \frac{3\pi}{2} \}$, so that the two amplitudes $x$ and $y$ must be $90$ degrees out of phase.

Choosing $x=\frac{1}{\sqrt{2}}$ we have that $y=\frac{i}{\sqrt{2}}$, and the transition matrices become
\begin{equation}
P = \frac{1}{\sqrt{2}}
\begin{pmatrix}
1 & i \\
0 & 0 \end{pmatrix}
\end{equation}
and
\begin{equation}
Q = \frac{1}{\sqrt{2}}
\begin{pmatrix}
0 & 0 \\
i & 1 \end{pmatrix}
\end{equation}
This results in Feynman's convention that each transition from \move{P} to \move{Q} or from \move{Q} to \move{P}, which are collectively referred to as zig-zags, corners or reversals, introduces a factor of $i$ to the amplitude.  Note that one obtains the same probabilities, and thus the same predictions, if $x$ and $y$ were rotated together by an arbitrary angle.

\subsection{Feynman Checkerboard}

The free particle illustrated in Figure \ref{fig:free-particle} can be interpreted by the observers in the space-time picture as zig-zagging along a chessboard using bishop-moves.  However, the fact that the two observers cannot order their mutual observations results in the fact that the particle can be thought of as simultaneously taking several zig-zag paths from the initial position in 1+1 dimensional space-time to a final position.  This is precisely the Feynman checkerboard model, which was introduced as an exercise in Feynman and Hibbs \cite{Feynman&Hibbs}.  In this exercise, it is noted that by assigning an amplitude proportional to $i \epsilon$ to each reversal in the zig-zag path, one obtains the Dirac equation in the limit that $\epsilon$ goes to zero.  In the present work we have taken this further by deriving both Feynman's model and Feynman's amplitude assignment from a simple model of a particle exhibiting direct particle-particle influence on two observers.

\section{Conclusion}

This paper summarizes the synthesis of three distinct research threads focused on probability theory, relativity and quantum mechanics, which were each inspired by Cox's derivation of probability theory via quantification of Boolean algebra. We show that these theories (derived from fundamental symmetries) when applied together enable one to make inferences about a simple model of direct particle-particle interaction give rise to well-known properties unique to Fermion physics, such as Zitterbewegung, uncertainty in position and momentum, as well as the Dirac equation in 1+1 dimensions.  Interesting work by Bialynicki-Birula \cite{Bialynicki-Birula:1994} suggests how one might extend this model naturally to 3+1 dimensions \cite{Earle:DiracMaster2011}.

Not only are the laws of inference derivable, but we are now finding that some laws of physics are derivable as well.  Fundamental symmetries originating from simple ordering relations are seen to constrain consistent apt quantification with the resulting constraint equations identified as physical laws.  In this sense, order literally gives rise to laws \cite{Knuth:laws}---which should not be surprising at all.

%%%%%%%%%%%%%%%%%%%%%%%%%%%%%%%%%%%%%%%%%%%%%%%%
%% BACKMATTER
%%%%%%%%%%%%%%%%%%%%%%%%%%%%%%%%%%%%%%%%%%%%%%%%

\section*{Acknowledgements}
The author would like to thank Keith Earle for many insightful discussions that have greatly influenced this work, as well as Ariel Caticha, Seth Chaiken, Philip Goyal, and John Skilling.



\bibliographystyle{aipprocl} % if natbib is missing

\bibliography{knuth}

\hyphenation{Post-Script Sprin-ger}
\begin{thebibliography}{10}
\providecommand{\enquote}[1]{``#1''}
\expandafter\ifx\csname url\endcsname\relax
  \def\url#1{\texttt{#1}}\fi
\expandafter\ifx\csname urlprefix\endcsname\relax\def\urlprefix{URL }\fi

\bibitem{Cox:1946}
R.~T. Cox, \emph{Am. J. Physics} \textbf{14}, 1--13 (1946).

\bibitem{Knuth:measuring}
K.~H. Knuth, \enquote{Measuring on lattices,} in \emph{Bayesian Inference and
  Maximum Entropy Methods in Science and Engineering, Oxford, MS, USA, 2009},
  edited by P.~Goggans, and C.-Y. Chan, American Institute of Physics, New
  York, 2009, AIP Conference Proceedings 1193, pp. 132--144, arXiv:0909.3684v1
  [math.GM].

\bibitem{Knuth&Skilling:2012}
K.~H. Knuth, and J.~Skilling, \emph{Axioms} \textbf{1}, 38--73 (2012),
  arXiv:1008.4831v2 [math.PR].

\bibitem{GKS:PRA}
P.~Goyal, K.~H. Knuth, and J.~Skilling, \emph{Phys. Rev. A} \textbf{81}, 022109
  (2010), arXiv:0907.0909 [quant-ph].

\bibitem{GK:Symmetry}
P.~Goyal, and K.~H. Knuth, \emph{Symmetry} \textbf{3}, 171--206 (2011).

\bibitem{Feynman:1948}
R.~P. Feynman, \emph{Rev. Mod. Phys.} \textbf{20}, 367--387 (1948).

\bibitem{Knuth+Bahreyni:EventPhysics}
K.~H. Knuth, and N.~Bahreyni, The physics of events: A potential foundation for
  emergent space-time (2012), arXiv:1209.0881 [math-ph].

\bibitem{Feynman&Hibbs}
R.~P. Feynman, and A.~R. Hibbs, \emph{Quantum Mechanics and Path Integrals},
  McGraw-Hill, New York, 1965.

\bibitem{Bialynicki-Birula:1994}
I.~Bialynicki-Birula, \emph{Phys Rev D} \textbf{49}, 6920--6927 (1994).

\bibitem{Earle:DiracMaster2011}
K.~A. Earle, A master equation approach to the `3 + 1' {D}irac equation (2011),
  arXiv:1102.1200 [math-ph].

\bibitem{Knuth:laws}
K.~H. Knuth, \enquote{Deriving laws from ordering relations.,} in
  \emph{Bayesian Inference and Maximum Entropy Methods in Science and
  Engineering, Jackson Hole WY, USA, August 2003}, edited by G.~J. Erickson,
  and Y.~Zhai, American Institute of Physics, New York, 2004, AIP Conference
  Proceedings 707, pp. 204--235, arXiv:physics/0403031v1 [physics.data-an].

\end{thebibliography}

\end{document}